\documentstyle[11pt,newpasp,twoside,epsf]{article}
\begin{document}
\title{The outer ejecta of $\eta$ Carinae}
\author{Kerstin Weis}
\affil{Institut f\"ur Theoretische Astrophysik, Universit\"at Heidelberg, 
Tiergartenstr. 15, 69121 Heidelberg, Germany}

\begin{abstract}
The nebula around $\eta$ Carinae consists of an inner bipolar
structure, historically called the {\it Homunculus}, and the outer
ejecta consisting mainly of a variety of knots of different sizes.
They reach out to distances of up to 30$^{\prime\prime}$ or
0.3\,pc from $\eta$ Car. With high-resolution long-slit
observations we mapped the outer nebula in order to analyze the
global expansion pattern and to model the three dimensional
structure of the ejecta. We find very different expansion
velocities for the knots of the outer ejecta. In some cases they
reach up to values as high as 2000\,km\,s$^{-1}$. Typical
expansion velocities lie at considerably lower values around $400 -
600$\,km\,s$^{-1}$, i.e., they are comparable to the expansion
velocities found in the Homunculus. Remarkably, the expansion of
the outer ejecta reveals a bidirectional motion pattern, which is
consistent with the bipolar structure of the inner nebula. A
general overview of the morphology and kinematics of the outer
ejecta is given and put into context with the structure and
kinematics of the inner part of the nebula, the Homunculus.  
\end{abstract}

\section{Introduction and historical background}

\begin{figure}[t]
\plotone{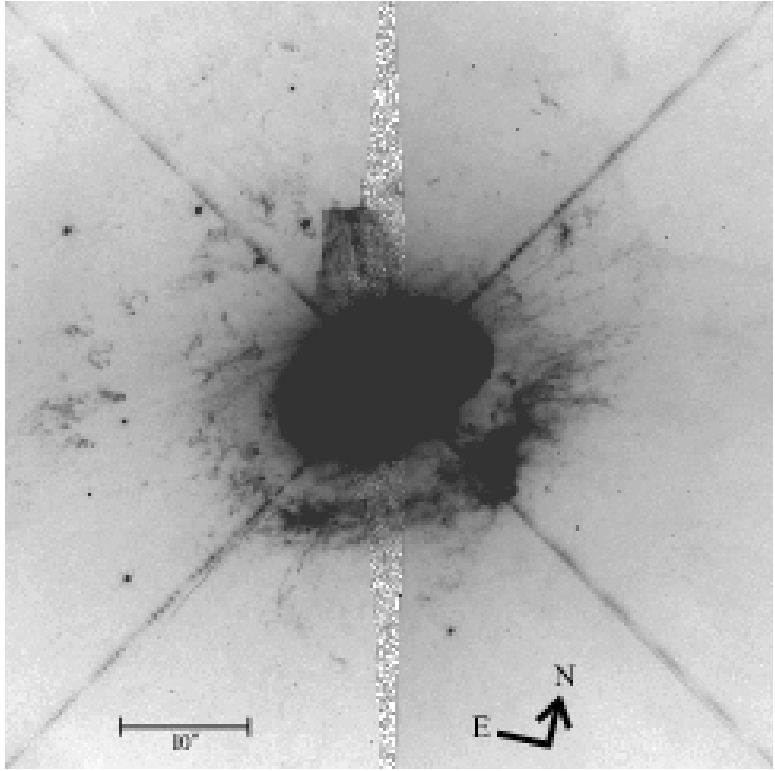}
\caption{This figure shows a deep HST image in the F656N filter, about
  60\arcsec\,$\times$\,60\arcsec\ large. The contrast was optimized to 
emphasize especially the faint emission of the outer ejecta. The inner
bipolar Homunculus appears as a more oval looking saturated structure in
the center.}
\label{fig:hstlong}
\end{figure}

$\eta$ Carinae has a bolometric luminosity of 
$L = 10^{6.7}$\,L$_{\sun}$  which puts it among the most 
luminous stellar objects known (Humphreys \& Davidson 1994; 
Davidson \& Humphreys 1997). In the light of recent discussions about $\eta$
Car being a binary (Damineli 1996; Damineli, Conti, \& Lopes 1997; 
Stahl \& Damineli 1998), it is noteworthy that the 
masses of the two components  would still be
between 65 and 70\,M$_{\sun}$ each. Both components would
therefore still be among the most massive stars. Historical
records already show the variable behavior of $\eta$ Carinae 
which was strongest during the star's giant eruption around 1843.
At that time $\eta$ Car reached $-1^{\rm m}$, making it the second
brightest star in the southern hemisphere at that time.
After that event the visual luminosity declined steadily down to 
$8^{\rm m}$ within a few years. Starting around 1950 the brightness
increased again and still does so today (for lightcurves see van Genderen 
\& Th\'e 1984; Viotti 1995; Humphreys, Davidson, \& Smith 1999 
and references therein). During the eruption of 1843
at least parts, if not all, of the nebula around $\eta$ Car 
formed. The nebula was not found until 
in 1938 van den Bos measured several components of the `$\eta$ Argus 
system' and noted that these `companions may be nebular nuclei'.
Only years later Gaviola (1946; 1950) and Thackeray (1949; 1950) detected
independently a nebulosity around $\eta$ Carinae. Due to its odd and 
`little man-like' shape, Gaviola named the nebula the 
{\it Homunculus}. Gaviola and Thackeray speculated already about a fuzzy 
and faint outer nebula or oval shell. 
With deeper images Walborn (1976) showed that the
nebula around $\eta$ Carinae was indeed much larger and consisted not 
only of the inner Homunculus.
He identified several knots outside this structure and named them according to
their positions and morphology. Examples are the {\it S ridge} and the 
{\it E condensations}. Ten years later, in 1985, using a
{\it shift and add} method an image of the Homunculus was made adding up 
200 short (0.25s) exposures. This image showed for the first time
the bipolar nature of the Homunculus (Duschl et al. 1995): two lobes,
one tilted towards the north-west, the other to the south-east.
The lobes are separated by an equatorial disk, which is defined 
through the so-called {\it streamers}.
With the launch of the {\it Hubble Space Telescope} (HST) the 
imaging capabilities 
improved even further and the first high-resolution image of the nebula around 
$\eta$ Carinae of the HST supported the model of a bipolar Homunculus.
Fig. \ref{fig:hstlong} shows an H$_{\alpha}$ (F656N-filter) HST image 
of the Homunculus and the outer structures. Note that the F656N filter 
does not only detect H$_{\alpha}$ emission but also [N\,{\sc ii}] emission,
since parts of the nebula show a large Doppler shift (see 
section \ref{section:kinematics}). In the following we will call
only the inner bipolar structure the Homunculus,  
all features and knots outside do form the {\it outer ejecta},
including the structures defined by Walborn (1976).  

\setlength{\textwidth}{4in}
\setlength{\textwidth}{5.25in}

Already Gaviola recognized a motion within the nebula comparing his 
images with older images in which individual parts of the nebula were
recorded as additional stellar components. Identifying these components 
with brighter parts of the Homunculus he obtained an expansion of several 
arcseconds per century or about 550\,km\,s$^{-1}$. A first detailed analysis 
of the expansion
of the knots was compiled in Ringuelet (1958) and Gratton (1963). Here
the measured expansion velocities again reach values of 
about 500\,km\,s$^{-1}$ if 
corrected for the distance, which is today believed to be around 2.3\,kpc
(Davidson \& Humphreys 1997; Walborn 1995 and references therein).
Using a longer time baseline ($10-15$ years) 
Walborn (1976), Walborn, Blanco, \&
Thackeray (1978), and Walborn \& Blanco (1988) determined 
proper motions of the outer
knots that range between 280\,km\,s$^{-1}$ and 1360\,km\,s$^{-1}$.
Modern measurements of the proper motion of the Homunculus are based 
on high-resolution HST images (Currie et al. 1996; Currie \& Dowling 1999;
Currie these proceedings). The velocities derived with this
method range between $10-1000$\,km\,s$^{-1}$.
An attempt to use old scanned images and resolution-degraded HST images 
was made by Smith \& Gehrz (1998).\\

One of the earliest observations to obtain radial 
velocities from the Homunculus 
dates back to Thackeray (1961) who measured velocities $\sim$ 850\,km\,s$^{-1}$
for the permitted lines and 220\,km\,s$^{-1}$ for the forbidden lines.
Assuming that the permitted lines show a Doppler shifted plus a scattered
component he concluded that the expansion velocity of the Homunculus 
should be about 630\,km\,s$^{-1}$. Newer long-slit 
spectra support these measurements. An extensive study of the radial velocity
of the Homunculus and the outer ejecta was carried out by Meaburn (1994) 
and Meaburn et al. (1993; 1996 and references therein). The radial 
velocities derived reach
from 250\,km\,s$^{-1}$ to at least $-$1200\,km\,s$^{-1}$. Fabry-Perot  
observations by Walborn et al. (1991) yield similar results.

\section{The new dataset and observations}

In order to analyze the morphology and 
kinematics of especially the outer ejecta around
$\eta$ Carinae we obtained high-resolution long-slit echelle spectra with
the 4\,m telescope at the Cerro Tololo Interamerican Observatory. 
For order selection a post-slit H$_{\alpha}$ filter was used and the cross
disperser was replaced by a flat mirror. With the 79 l/mm 
grating the spectral resolution was about 14\,km\,s$^{-1}$ at 
the H$_{\alpha}$ line.
The slit length was vignetted to 4\arcmin, the relevant section was 
only 1\farcm5.  The pixel size in the spatial direction was 
0\farcs264 per pixel.
The observations were performed at a position angle (PA) of 132\deg, 
i.e., parallel to the major axis of the Homunculus.
Starting at an offset star (star \#60 in Th\'e, Bakker, \& Antalova 1980) 
the spectra 
were offset 2\arcsec\ to the north or south. {\it Slit 12S} 
is therefore a slit, offset by 12\arcsec\ to the south of the offset star.
The positions of all slits are depicted in Fig.\,\ref{fig:slits} Note that 
due to the PA of 132\deg\ the 2\arcsec\ offsets translate into a spacing 
between parallel slits of only 1\farcs5. In all, 31 spectra were taken,
of which the central 6 slits could 
not be used because of strong stray-light from the central star.

\setlength{\textwidth}{4in}
\begin{figure}[t]
\plotone{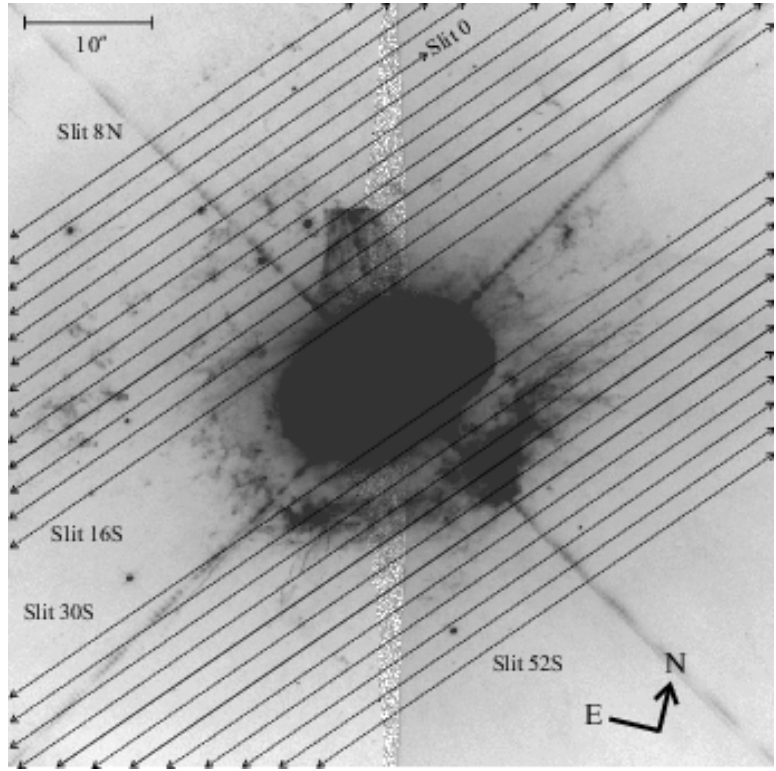}
\caption{Here the same HST image as in figure \ref{fig:hstlong} is shown, 
  with our slit
  positions overplotted. The central 6 slit
  positions crossing the Homunculus are missing, since they could not be used
  due to strong stray-light. Solid lines indicate slits for which
  echellograms are depicted in figure \ref{fig:echelle}.}
\end{figure}
\label{fig:slits}
\setlength{\textwidth}{5.25in}

\section{The morphology of the outer ejecta}

Using deep HST images taken with the F656N and F658N filter the 
outer ejecta was analyzed and its morphology studied.
In contrast to the inner bipolar Homunculus the outer ejecta shows no 
clearly symmetric structure.
A large amount of knots, bullets and filaments surround $\eta$ Carinae 
up to a distance of 30\arcsec\ in radius (0.33\,pc), compared to the inner 
Homunculus which has a diameter of 17\arcsec\ (about 0.2\,pc). The sizes 
and shapes of the knots and filaments in the outer  ejecta are manifold, 
reaching between 7\farcs5 for the N condensation to fractions of arcseconds
for the smallest structures. Most likely many small filaments are not
resolved yet, and very faint, small structures are below the detection 
limit. The morphology of the structures in the outer ejecta represent 
those of irregular shaped knots and bullets, some might even be 
described as arches, some look more like long filaments. 
Most amazing are very straight, long collimated structures 
which in the following will be called {\it strings}, these structures are 
discussed in section \ref{section:strings} in more detail. 
Other than these individual structures  the {\it South ridge} is the most
remarkable feature in the outer ejecta. Here a large amount of 
bullets seem to be 
connected or even more likely are running into each other. As it will be 
shown later these colliding knots are giving rise to X-ray emission 
from this area (see section \ref{section:xray} and Weis, 
Duschl, \& Bomans 2000).

\section{The kinematics of the outer ejecta}\label{section:kinematics}

\begin{figure}[t]
\plotone{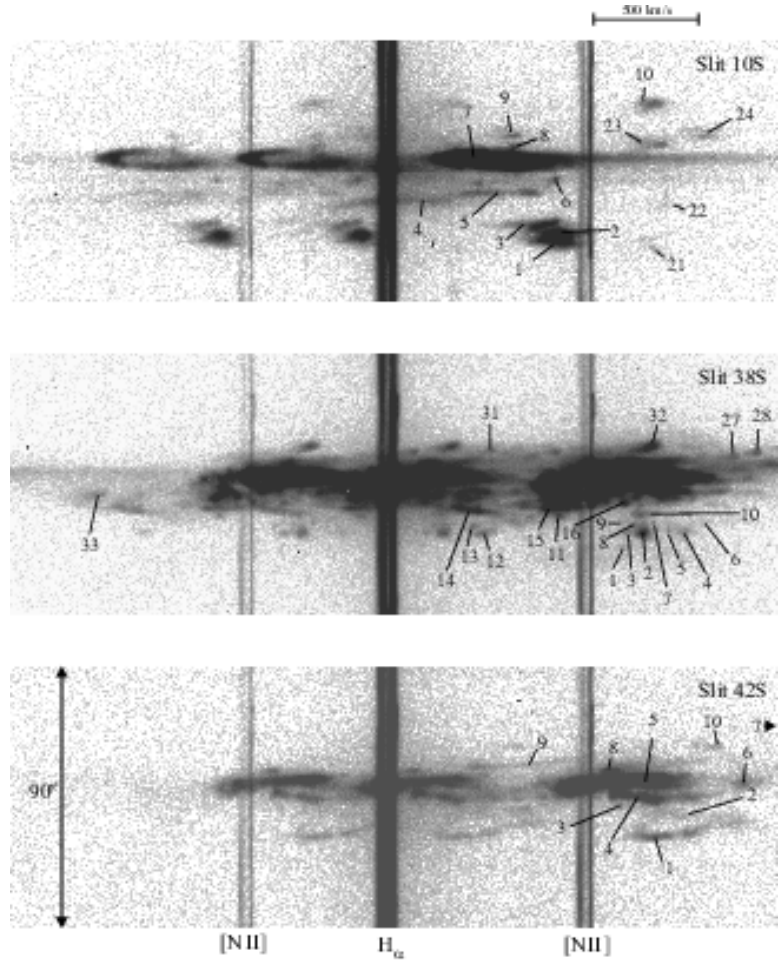}
\caption{The 3 echellograms depicted here are 75\,\AA\ wide and 90\arcsec\
  high. The positions of the slits are marked with solid lines in figure 
\ref{fig:slits} On top of the echellograms a bar indicates a Doppler shift of
500\,km\,s$^{-1}$.}
\label{fig:echelle}
\end{figure}

The kinematics of the outer ejecta was studies using the high-resolution 
long-slit echelle data. For each knot detected in the spectra the minimum and 
maximum velocities were derived, as well as the velocity of its brightest 
emission. Fig.\,\ref{fig:echelle} shows 3 typical echellograms. 
Each one is 75\,\AA\ wide, centered on the 
H$_{\alpha}$ line at rest, and a spatial section of  90\arcsec\ is  shown. 
In the echellogram red and blueshifted knots are visible and
labeled by numbers. A single knot is  a coherent structure, which  
may or may not consist out of several sub-entities visible due to 
their different intensities or velocities, see for example 
knot\,7 in Slit 10S.  
Taking  all measurements of the spectra together the following 
results were deduced:
\begin{itemize}
\item{ About 200 knots were identified 
from which the fastest redshifted feature 
moves with $+2000$\,km\,s$^{-1}$ while the 
fastest blueshifted structures' velocity 
is $-1200$\,km\,s$^{-1}$.}
\item{ Most of the knots move slower but still with $+600$\,km\,s$^{-1}$ to 
$-600$\,km\,s$^{-1}$. Note that these values are only lower limits of the 
true expansion velocities since we measure radial velocities, which do not 
take into account the projection angle. If it were 45\deg\ this 
would yield already velocities around 850\,km\,s$^{-1}$ for most of the knots.}
\item{ The largest spread in velocity of a single coherent knot was found to 
be 1250\,km\,s$^{-1}$.}
\end{itemize}

\setlength{\textwidth}{4in}
\setlength{\textwidth}{5.25in}

\begin{figure}
\plotone{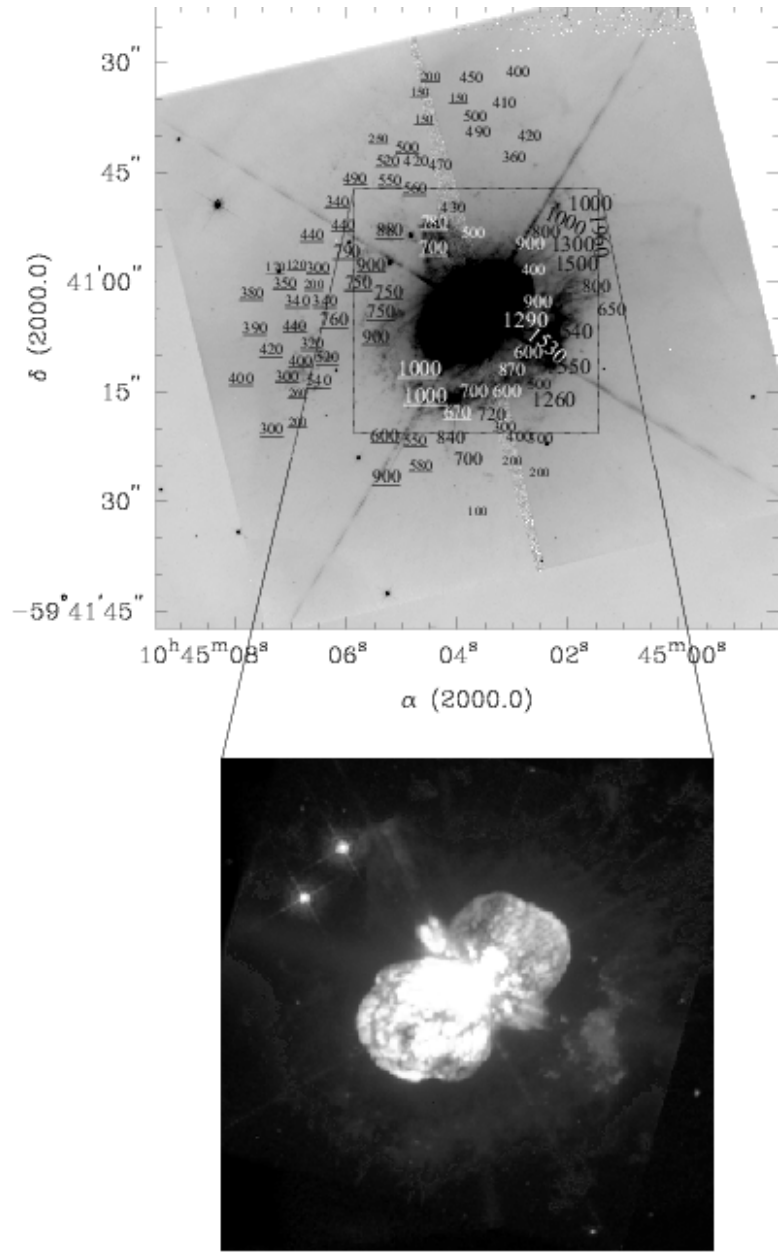}
\caption{In the upper panel the same image as in Fig.\,\ref{fig:hstlong} 
is shown. Onto the image, scaled by font sizes, the radial 
velocities of individual knots (with the highest velocity for 
a certain area) are overplotted.
Underlined velocities represent negative (blushifted) structures,
not underlined ones are redshifted, positive values. A clear trend is visible,
showing that blueshifted structures appear in the south-east while redshifted
are concentrated to the north-west. The lower panel shows the central 
bipolar Homunculus for comparison.}
\label{fig:bipolar}
\end{figure}

Beside these results for individual knots also a global trend of the 
expansion in the outer ejecta was discovered. Fig.\,\ref{fig:bipolar} 
visualizes this result. In this figure the velocities of representive 
knots are overplotted onto an HST image. Blueshifted, negative velocities 
are underlined. A systematic distribution is then obvious: Knots with 
redshifted emission are 
concentrated towards the 
north-west of the star, while knots in the south-east show 
predominantly blueshifted lines. With respect to the central star, 
this represents a bidirectional motion pattern, i.e., also the outer 
ejecta shows a bipolar structure. If we compare the outer ejecta 
with the inner 
bipolar Homunculus, we can see a similarity. The north-western lobe 
of the Homunculus is tilted away from the observer, its net expansion 
is redshifted, while the south-eastern lobe  is blueshifted, 
this lobe is tilted towards us. Therefore the Homunculus and the outer ejecta
show not only both a bipolar morphology, but one with a very similar
orientation. 

In the echelle spectra also the [N\,{\sc ii}]$\lambda$6583\AA/H$_{\alpha}$ 
ratio could be measured, a rough 
indicator for the nitrogen overabundance of the ejecta. For knots in the 
outer ejecta [N\,{\sc ii}]$\lambda$6583\AA/H$_{\alpha}$ ratios 
between 1 and 5 were measured, with most of them 
around 3.

\setlength{\textwidth}{4in}
\setlength{\textwidth}{5.25in}

\section{Strings}\label{section:strings}

\begin{figure}
{\plottwo{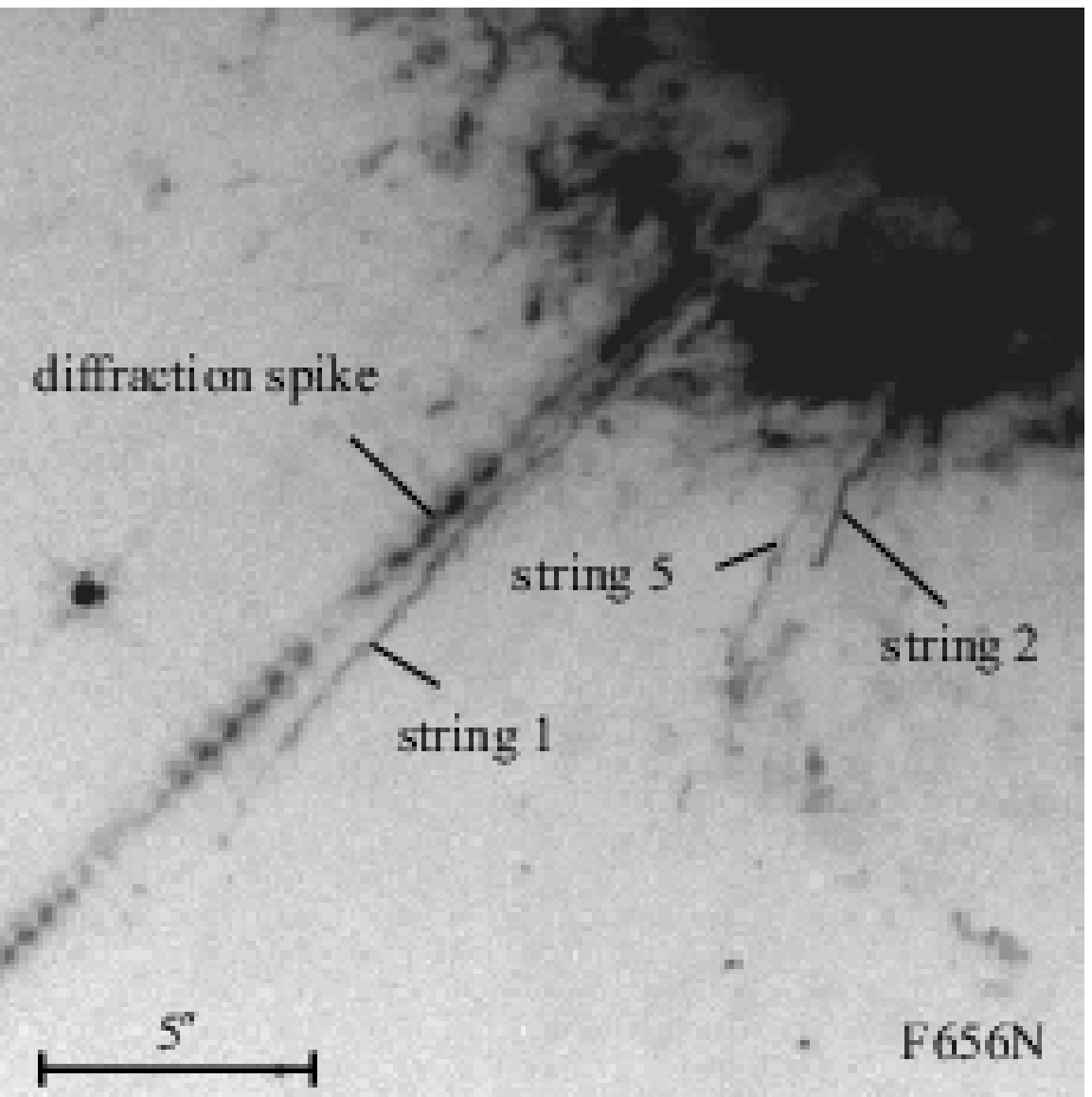}{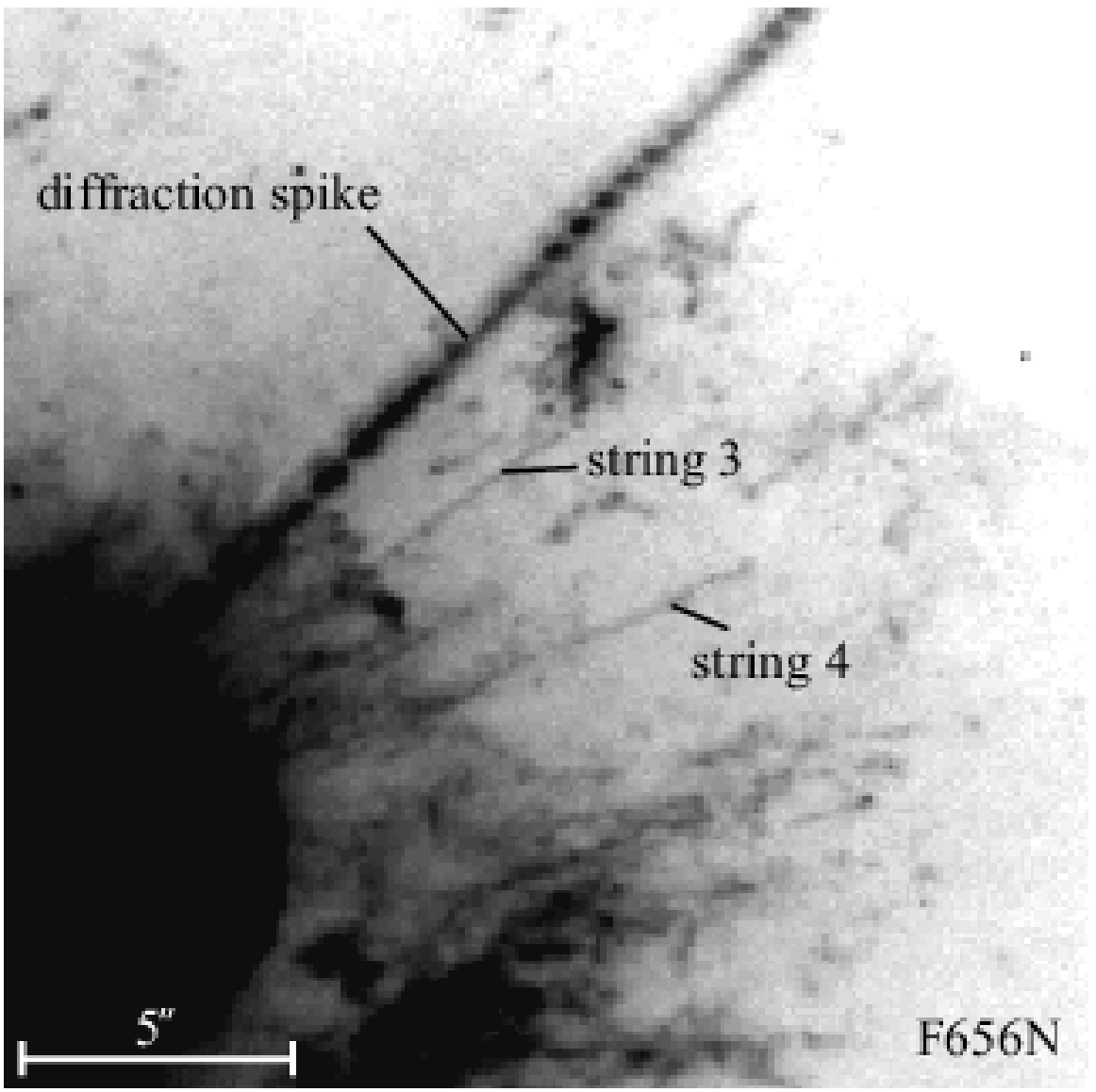}}
\caption{{\it Left:} Enlargement of the F656N HST image
showing the strings 1, 2 and 5 in detail. 
{\it Right:} To the north-east of $\eta$ Carinae,
strings 3 and 4 are visible.}
\label{fig:strings}
\end{figure}

In the outer ejecta, we found very straight, long and highly collimated 
structures the strings. They are of great interest in relation to 
the formation mechanism of the nebula around $\eta$ Carinae.
We identified 5 such strings in the nebula by visual inspection 
of the HST images. On much smaller scales numerous such 
structures can be found, but we do not count them among the strings 
since their collimation and parameters are not as extreme as those of the 
strings.  With a length of 0.18\,pc and a length-to-width 
ratio of 70 String\,1 is the largest string and comparable in 
size to the entire Homunculus nebula.
Images of the strings are shown in Fig. \ref{fig:strings}.
In addition to their amazing morphological structure  the 
strings show a strange 
kinematic behavior. Their radial velocity increases
along the string outwards. Close 
to the star, String\,1 moves with  $-522$ \,km\,s$^{-1}$, 
following the string outwards this velocity increases steadily  up to 
$ \sim -1000$\,km\,s$^{-1}$ 
at the far end. The strings follow an almost perfectly linear velocity law. 
Extrapolating the strings back to the star, this relation makes it exceedingly
likely that 
the radial velocity reaches zero at the star. The strings show a 
Hubble type velocity law. A more detailed analysis 
of the strings can be found in Weis, Duschl, \& Chu (1999).
Still, it is not clear what the physical nature of the strings is.
They may well be highly collimated coherent structures, 
similar to a water jet, for instance, but one may equally well 
envisage a train of many individual knots or bullets just
following the same path. 
Another possibility is that they are trails or wakes 
following an object at the strings' far ends, or even projection 
effects of the walls of, for instance, much wider funnels. 
The origin and physics behind the strings is not solved yet.
Recently, we have obtained HST-STIS data from which we expect to be able to
conclude on their physical status (densities but also physical structure).

\section{The X-ray emission from the outer ejecta}\label{section:xray}

X-ray images of the nebula around $\eta$ Carinae (Fig.\ \ref{fig:xray},
ROSAT HRI image overlayed as contours over an optical F656N filtered HST 
image) show no clear symmetry or coincidence with the optical emission.
The X-ray 
emission is hook-shaped with two brighter maxima---one of which appears 
close to the S ridge. When comparing the X-ray emission 
not only with the optical 
emission but with the kinematics of the outer ejecta a much better 
correlation is found. 
X-rays are present where fast knots are located. The emission 
is due to fast knots producing shock waves. These 
shocks are presumably strongest where the density is highest.
This is for instance in agreement with the bright X-ray spot being seen at
the S condensation's location where many knots are indicative of higher 
density.

\setlength{\textwidth}{4in}
\begin{figure}
\plotone{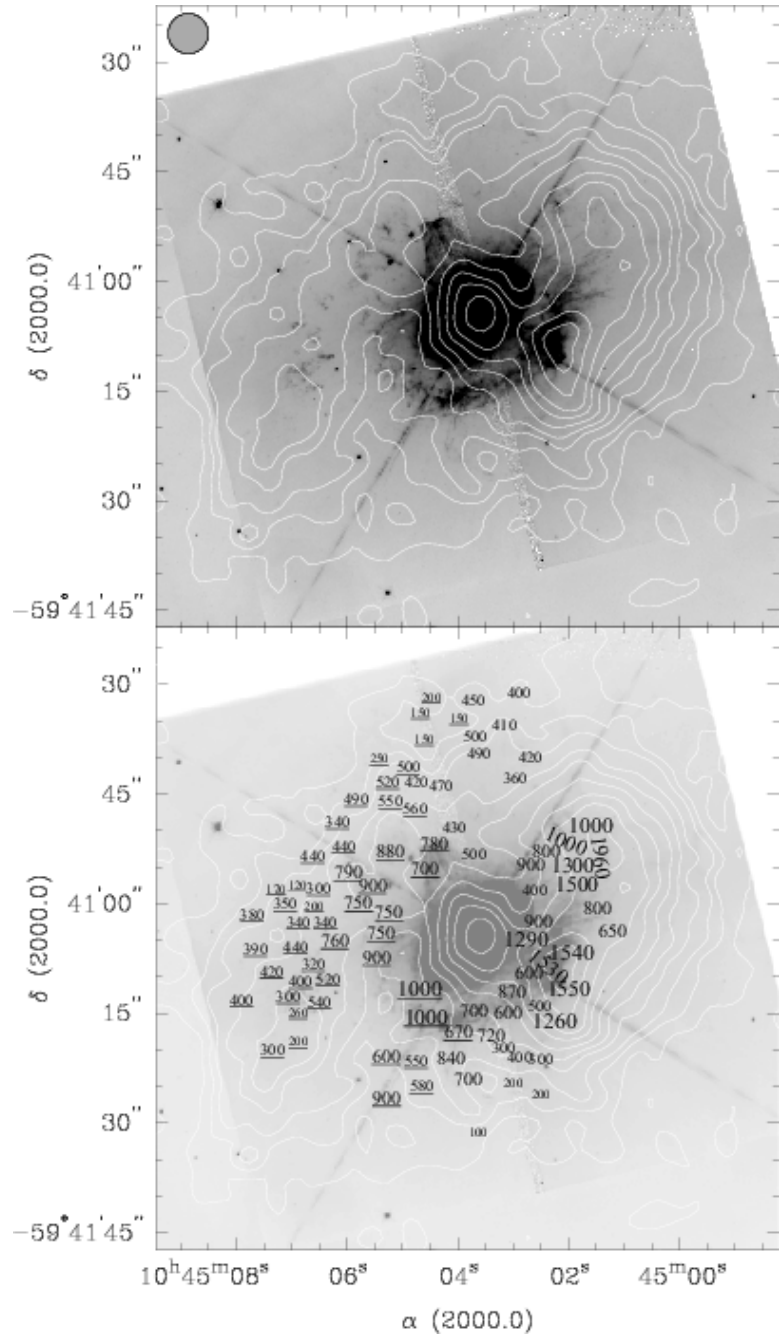}
\caption{Overlay of the X-ray emission (contour lines) and
an HST image (grey scales). The lower
panel shows the same overlay as the upper panel (with decreased grey scale
intensity), with the measured radial velocities (in km\,s$^{-1}$) placed
at their respective positions. Underlined numbers indicate negative
velocities. The sizes of the characters increase with increasing
absolute velocities (see text).}
\end{figure}
\label{fig:xray}
\setlength{\textwidth}{5.25in}

Since the majority of the knots in the outer ejecta shows velocities around
400\,km\,s$^{-1}$ to 600\,km\,s$^{-1}$ we derive a 
plasma temperature of $2-5\,10^6$\,K,
using $T_{\rm ps} = V^2 ( 3 \mu / 16 k)$ (see McKee 1987; $V$: radial 
velocity taken as
representative for the shock velocity; $\mu$: mean mass per
particle). We assumed a normal He/H ration of
0.1 leading to $\mu=0.61$ for a fully ionized gas. Since the material in
the nebula around $\eta$ Car contains a significant amout of CNO processed
material the He/H ratio is most likely higher, which also increases the 
plasma temperature (by about a factor of 1.5 for He/H$=0.33$). 
The highest X-ray emission results from two X-ray knots where the 
velocities measured from the spectra are much higher than average. 
Here the radial velocities 
($+1500$ and $+1900$\,km\,s$^{-1}$) indicate equilibrium shock temperatures 
of  $3\,10^7$ and $5\,10^7$ K. 
For further details of this analysis see Weis, Duschl, \& Bomans (2000).

\section{Summary and Conclusions}

The outer ejecta is in many respects an outstanding nebula, which
extends to about 0.33\,pc in radius around $\eta$ Carinae.
The measured expansion 
velocities  reach as high as  2000\,km\,s$^{-1}$, i.e., much higher than 
expected and higher than detected earlier. The majority of the knots in 
the outer ejecta as well as the  Homunculus itself move with 
velocities between $+600$ and $-600$\,km\,s$^{-1}$. While the 
Homunculus shows a bipolar 
morphology, the knots in the outer ejecta seem randomly distributed 
and do not obviously follow a certain symmetry. However, taking the 
kinematic data into account, the outer ejecta {\it does} show a 
bipolar symmetry indicated by the distribution of blue and redshifted 
structures. The ejecta itself resembles 
a large collection of bullets of very different sizes and
morphologies. 
Moving with higher velocities these knots form shocks which are then 
visible in X-rays. A comparison of the morphology of the X-ray  emission
with the kinematics of knots in the outer ejecta nicely demonstrates
this coincidence. \\

\acknowledgements

Sincere thanks goes to Prof.\ Wolfgang J.\ Duschl for reading and 
improving the manuscript and to Dr.\ Dominik J.\ Bomans for 
inspiring discussions on the subject. This work was supported by the 
Deutsche Forschungsgesellschaft (DFG) through grant Du\,168/8-1.

\end{document}